\documentclass[10pt,letterpaper]{article}
\usepackage[top=0.85in,left=2.75in,footskip=0.75in]{geometry}
\usepackage{changepage}
\usepackage{amsmath,amsthm}
\usepackage{amsfonts,amssymb}
\usepackage{enumitem}
\usepackage[utf8x]{inputenc}
\usepackage{textcomp,marvosym}
\usepackage[right]{lineno}
\usepackage{microtype}
\DisableLigatures[f]{encoding = *, family = * }
\usepackage{array}
\usepackage{setspace}
\usepackage{advdate}
\usepackage{url}
\usepackage{soul}
\usepackage{slashbox}
\usepackage{hyperref}
\usepackage{cuted, nccmath}
\usepackage{bm, bbm}

\usepackage[table]{xcolor}
\usepackage{graphicx}
\newtheorem{theorem}{Theorem}
\usepackage{cite}
\DeclareMathOperator{\halpha}{\hat{\alpha}}
\DeclareMathOperator{\hbeta}{\hat{\beta}}
\DeclareMathOperator{\ind}{\mathbbm{1}}

\newcommand{\red}[1]{{\textcolor{black}{#1}}}

\newcommand{\changeR}[1]{{\textcolor{black}{#1}}}

\date{}
\newcolumntype{+}{!{\vrule width 2pt}}

\newlength\savedwidth

\raggedright
\usepackage[aboveskip=1pt,labelfont=bf,labelsep=period,justification=raggedright,singlelinecheck=off]{caption}

\makeatletter
\renewcommand{\@biblabel}[1]{\quad#1.}
\makeatother

\usepackage{lastpage,fancyhdr,graphicx}
\usepackage{epstopdf}
\fancyheadoffset[L]{2.25in}
\fancyfootoffset[L]{2.25in}
\begin{document}

\vspace*{0.2in}

\begin{flushleft}
{\Large
\textbf\newline{Directionally dependent multi-view clustering using copula model} 
}
\newline
\\
Kahkashan Afrin\textsuperscript{1\ddag},
Ashif S. Iquebal\textsuperscript{1\ddag},
Mostafa Karimi\textsuperscript{2\ddag},
Allyson Larsen\textsuperscript{3\ddag},
Se Yoon Lee\textsuperscript{3\ddag},
Bani K. Mallick\textsuperscript{3\ddag*}
\\
\bigskip
\textbf{1} Department of Industrial \& Systems Engineering, Texas A\&M University, College Station, TX, USA
\\
\textbf{2} Department of Electrical and Computer Engineering, Texas A\&M University, College Station, TX, USA
\\
\textbf{3} Department of Statistics, Texas A\&M University, College Station, TX, USA
\\
\bigskip
\ddag These authors contributed equally to this work.

* bmallick@stat.tamu.edu

\end{flushleft}
\section*{Abstract}     

\changeR{Recent developments in high-throughput methods have resulted in the collection of high-dimensional data types from multiple sources and technologies} that measure distinct yet complementary information. Integrated clustering of such multiple data types or multi-view clustering is critical for revealing pathological insights. However, multi-view clustering is challenging due to the complex dependence structure between multiple data types, including directional dependency. \changeR{Specifically, genomics data types have pre-specified directional dependencies known as the central dogma that describes the process of information flow from DNA to messenger RNA (mRNA) and then from mRNA to protein}. Most of the existing multi-view clustering approaches assume an independent structure or pair-wise (non-directional) dependence between data types, thereby ignoring their directional relationship. Motivated by this, \changeR{we propose a biology-inspired Bayesian integrated multi-view clustering model that uses an asymmetric} copula to accommodate the directional dependencies between the data types. \changeR{Via extensive simulation experiments, we demonstrate the negative impact of ignoring directional dependency on clustering performance}. \changeR{We also present an application of} our model to a real-world dataset of breast cancer tumor samples collected from The Cancer Genome Altas program and \changeR{provide comparative results.} 



\section*{Introduction}
\changeR{The advancements in high throughput technologies and the emergence of several supportive programs such as The Genome Technology Program at the National Human Genome Research Institute and The Cancer Genome Atlas program have enabled the capabilities for rapid, high-quality, and low-cost collection of genomics data from multiple sources~\cite{GTP, cancer2012comprehensive}}. These data types, collected from several heterogeneous sources for the same set of objects or patients,  often provide unique but complementary information. They can be thought of as providing different views for the same underlying phenomenon (with each data type representing a particular view) and thus are referred to as ``multi-view" datasets. 

\changeR{With this explosion of data, there is a strong need for integrated analysis of multi-view data to not only provide an immense amount of added information for making inference about the objects but also to explore and utilize the complex associations between multiple data types~\cite{lock2013bayesian}.
Hence, an integrated analysis of multi-view data has emerged as one of the promising areas of research. In several studies, integration of multiple data types has been shown to provide more comprehensive and radically new perspectives for understanding molecular pathways and the progression of diseases such as cancer as compared to analyzing individual data types separately~\cite{lock2013bayesian,karczewski2018integrative, reif2004integrated}.}

In this research, we are concerned with proposing an effective approach for an integrated multi-view clustering where the objective is to group a set of patients, based on different genomic data types---an emerging field with only a few publications so far~\cite{ickstadt2018toward}. Following the existing literature in this domain, we refer to this integrative clustering as vertical multi-view clustering or consensus clustering \cite{lock2013bayesian,ickstadt2018toward, kirk2012bayesian}. 

Existing studies on vertical multi-view clustering can be primarily  categorized into the following two groups~\cite{lock2013bayesian}: 
\begin{enumerate}[leftmargin=0pt]
	\item[](i) Separate or source-specific clustering of each data type followed by post-hoc integration of the clustering outcomes often without incorporating any association between the data types \cite{wang2011bayesian,bruno2009multiview}. \changeR{This is also referred to as late integration or ensemble clustering \cite{rappoport2018multi}.}  
	    \item[](ii) \changeR{Concatenating the data types prior to clustering (or early integration) to obtain a single or unified model using the concatenated/joint data \cite{shen2009integrative}.} 
\end{enumerate}

\changeR{The two-stage late integration approach often fails to explore and exploit the association between different data types by assuming no dependence structure. On the other hand, early integration with concatenated data can have scaling and high-dimensionality issues and fail to recognize the individual contribution of each data type ~\cite{lock2013bayesian}. }
  
Hence, effective multi-view data integration methods are required that accommodates the dependence structure across the data types. Incorporating such dependencies between the multiple data types has been shown to encapsulate comprehensive information and deep understanding \cite{reif2004integrated, chaudhuri2009multi}. However, effective multi-view data integration to capture the dependence between multiple data types remains a key challenge~\cite{zhao2017multi}.

\changeR{Recently, authors in \cite{kirk2012bayesian} proposed an approach for integrative analysis via multiple dataset integration (MDI) by modeling each dataset (or data types) using a Dirichlet-multinomial mixture model and the association between the data types was captured by using the pairwise dependence between the clusters.} Their method allowed for the identification of groups of genes that often fell together in one cluster. \changeR{However, it did not provide a direct route to obtain the overall clustering, which is often of interest in practical applications.} Along the similar lines, authors in \cite{lock2013bayesian} proposed a statistical model, \changeR{called Bayesian consensus clustering (BCC)}, for integrating two or more data types. \changeR{Similar to MDI, BCC also assumes a Dirichlet-multinomial mixture model for the data types}. However, their approach was based on defining a source-specific cluster (i.e., separately clustering the objects for each of the datasets) as well as a consensus (i.e., overall) clustering. The dependence between the data types was captured by defining a parameter that controls the adherence of the source-specific clustering to the consensus clustering. They also emphasized on the computational scalability and robustness of the Bayesian framework for simultaneously estimating the consensus clustering as well as the source-specific clustering {as compared to both late and early integration~\cite{lock2013bayesian}.} 

\changeR{These recent studies on integrative analysis have focused both on the issue of source-specific and provides an effective way of consensus clustering while accounting for the dependence between multiple data types. However, the dependence between different data types is extremely complex and is governed by the underlying molecular biology (in the case of genomics data)}. Specifically, genomics data types have dependencies that are often directional. For example, the central dogma of molecular biology~\cite{crick1958protein,crick1970central} describes the flow of information from DNA to messenger RNA (mRNA) through transcription and then from mRNA to protein through translation \cite{alberts2017molecular,kim2017integrative}, making these data types directionally dependent. \changeR{Further, the central dogma explains that this transfer of information is acyclic (with a pre-specified direction), for example, the transfer of information from gene to protein and not from protein to protein or protein to gene \cite{weber2006central}}.

\changeR{This directional relationship between the data types is crucial for explaining physiological traits and clinical outcomes~\cite{qin2019identifying}. Furthermore, it has been analytically proven that the more remote an omics level or data source is from a physiological trait, the smaller the magnitude of their correlation is. For instance, the proteome-trait (protein level) correlation test is more powerful than the transcriptome-trait (RNA level) correlation test, which in turn is more powerful than the genotype-trait (DNA level) correlation test \cite{qin2019identifying}. Nonetheless, the current multi-view clustering studies in the literature do not incorporate this directional information into their modeling. To address this gap, we propose a biology-inspired integrated multi-view clustering model called Bayesian directional multi-view clustering that incorporates the directional dependence between the data types using a copula function}. 

Copulas are multivariate distribution functions that allow us to model the dependence structure by considering the marginals \cite{nelsen2007introduction}. Owing to the modeling flexibility provided by copulas, they have been used extensively in the literature for obtaining the dependencies between the data types. For instance, \cite{rey2012copula} used a Gaussian copula to construct a Dirichlet prior mixture of multivariate distributions to perform dependency-seeking clustering and showed significant improvement in the clustering results. Nonetheless, to the best of our knowledge, no existing work in the multi-view clustering has employed directional dependencies. 

In this work, we obtain the directional dependence between the data types using an asymmetric copula regression. The use of asymmetric copula is crucial for modeling the directionality as symmetric copulas can only provide the directional dependence in the marginal behaviour but not in the joint behavior, as pointed out by \cite{sungur2005note}. Therefore, symmetric copulas may not be used to capture directional dependence. Here, we used the Rodriguez-Lallena and Ubeda-Flores \cite{rodriguez2004new} family of asymmetric copulas to capture the directional dependence. Further, we analyzed the asymmetric copulas from a regression perspective that allows us to obtain not only the direction of dependence between the data types but also to quantify this directional dependence~\cite{sungur2005note}. 

To evaluate the efficacy, we applied our model to both synthetic as well as a \changeR{benchmark} real-world dataset. Using the results, we demonstrate that modeling the directional dependence between different datasets improves the clustering performance. For the real-world application, we used the dataset of breast cancer tumor samples that is publicly available from The Cancer Genome Atlas (TCGA) program~(\url{https://www.cancer.gov/tcga})~\cite{lock2013bayesian,cancer2012comprehensive}.

The rest of the article is organized as follows: In the \textbf{Methods} section, we describe the background on the Dirichlet mixture model for the integrative analysis and copula to capture directional dependence. The \textbf{Copula-based multi-view clustering} section presents our copula-based multi-view clustering approach and the posterior inference. We describe the simulation and case study examples along with the results and comparative analysis in the \textbf{Results} section. Finally, we conclude the paper with a brief discussion and future works in the \textbf{Conclusion} section.

\section*{Methods}
\changeR{In this section, we briefly introduce the Dirichlet mixture model and extend it to a multi-view setting. Next, we present an overview of copulas and discuss a copula framework for capturing the directional dependence between multiple data types.}   

\subsection*{Dirichlet mixture models} \label{section:mixture}
Let $\{\textbf{X}_m\}_{m=1}^{M}$ denote a collection of $M$ distinct data types for $N$ objects (e.g., patients with different cancer tumor), and for each data type $m$, the notation $X_{mi}$ represents the data corresponding to the $i$-th object. For example, if the first data source (indexed by $m=1$) is RNA gene expression, then $X_{1i}$ denotes the RNA gene expression profile for the $i$-th patient. \changeR{In the context of this work, we assume that each data type is available for all $N$ objects.} For each data type, $m$, let $L_{mi}=\{1,2,\ldots,K\}$ denote a latent variable corresponding to the data $X_{mi}$ \changeR{such that} $L_{mi} = k$ implies that the $i$-th object belongs to the $k$-th cluster. 

\changeR{Since the objective of multi-view clustering is to partition the $N$ objects into $K$ clusters using the integrated data, it is intuitive to consider that the data $X_{mi}$ for each data type $m$ is generated from a mixture density given as \cite{mclachlan1988mixture,bishop2006pattern}}:
\begin{equation}
\label{eq:mixture_density}
p_m(X_{mi}) = \sum_{k=1}^{K}\pi_{mk}f(X_{mi}|\theta_{mk})
\end{equation}
where $\pi_{mk} = \text{Pr}[L_{mi} = k] \in [0,1]$ represents the probability of $X_{mi}$ belonging to the $k$-th cluster, and $f(X_{mi}|\theta_{mk})$ is a probability density function for the data $X_{mi}$ indexed by the parameters $\theta_{mk}$. If $f(\cdot)$ is chosen to be a Gaussian density $N(\mu_{mk},\sigma_{mk}^2)$ with mean $\mu_{mk}$ and variance $\sigma_{mk}^2$, then we have $\theta_{mk} = (\mu_{mk},\sigma_{mk})$, leading to a Gaussian mixture model \cite{rasmussen2000infinite}. For a detailed explanation for the mixture model, refer to~\cite{bishop2006pattern}. With this, a hierarchical structure of Dirichlet mixture model for each data source $m$ may be obtained as \cite{walker2007sampling,hjort2010bayesian,muller2015bayesian}:
\begin{align}
\label{eq:CDF_F_in_Dirichlet mixture model}
X_{mi}|L_{mi},\bm{\theta}_m &\sim F(\theta_{m L_{mi}});\quad \theta_{mk} \sim G^{(0)}
\\
L_{mi}|\pi_m &\sim \text{Multinomial}(\pi_{m1},\pi_{m2},\ldots \pi_{mK})  \label{eq:MixtureModel}
\\
 \label{eq:Dirichlet}
\pi_{m1},\pi_{m2},\ldots \pi_{mK} &\sim \text{Dirichlet}(\alpha/K,\ldots.\alpha/K)
\end{align}
where $F$ is the (cumulative) distribution function for the density $f(\cdot)$ participating in the mixture density in Equation~(\ref{eq:mixture_density}), $G^{(0)}$ denotes a base measure for the parameter $\theta_{mk}$, and $\alpha>0$ denotes the scaling parameter for the Dirichlet distribution in Equation~(\ref{eq:Dirichlet}). For a detailed description for the Dirichlet mixture model, and its practical implementation for clustering, refer to \cite{gorur2010dirichlet}.

\subsection*{Directional dependency and copula} \label{section:Motivation to Accommodate Directional Dependency}
One of the fundamental ideas of molecular biology is the central dogma \cite{crick1958protein,crick1970central}, which describes the process of information flow via a two-step process of transcription and translation, where the information in genes flow to proteins: DNA to mRNA, and mRNA to protein. \changeR{Further, the central dogma explains that this transfer of information is acyclic (with a pre-specified direction), for example, the transfer of information is from gene to protein and not from protein to protein or protein to gene \cite{weber2006central}}.

The key motivation of our research is to subsume this \emph{directional dependency} into a multi-view clustering problem \cite{kim2017integrative,lock2013bayesian,wang2014breast}, leading to a unified and \changeR{directionally dependent} clustering model.

\changeR{Directional dependence was first presented by authors in \cite{dodge2000direction} using an ordinary linear regression model. Under this framework, a random variable $X$ is directionally dependent on $Y$, if the square of sample skewness of $Y$, denoted by $\hat{\gamma}^2_Y$, is less than that of $X$ when we regress $X$ on $Y$. That is, $\hat{\gamma}^2_Y < \hat{\gamma}^2_X$ when $X=\beta Y + \epsilon$. The fundamental notion here is that the square of the skewness of the response variable in a linear regression setting is always less than equal to the square of the skewness of the explanatory variable.} More recently, authors in \cite{sungur2005some,sungur2005note,kim2014analysis, jung2008new} argued that copula regression models might offer a possibility to capture the directional dependence between variables. This is mostly because the copula regression approach can model the joint dependence structure between the random variables, independently from the choice of the marginal distributions. 

In this paper, we make use of copulas \cite{trivedi2007copula,jaworski2010copula} to accommodate directional dependency into a multi-view clustering framework. Copulas are widely used to model the dependence structure between random variables by decoupling the dependence structure from the marginal distribution \cite{demarta2005t}. \changeR{Specifically, an $n$-dimensional copula function $C = C(u_1,u_2,\ldots, u_n)$ is a multivariate distribution function defined on a unit hypercube with uniform marginals given as:
\begin{equation}
C(u_1,u_2,\ldots, u_n) = P(U_1\leq u_1,\ldots, U_n\leq u_n)
\end{equation}where $U_i \sim\text{uniform}(0,1), i=1,2,\ldots,n$. Given a vector of random variables denoted by $\mathbf{X}=\{X_1,X_2, \ldots, X_n\}$ with joint distribution function given as: \[F(x_1,x_2,\ldots,x_n) = P(X_1\leq x_1, X_2 \leq x_2, \ldots, X_n \leq x_n),\] the uniqueness of the copula associated with $F$ was initially observed in \emph{Sklar's theorem} \cite{sklar1959fonctions}.}
\begin{theorem}[Sklar's theorem]
	Let $F$ be an $n$-dimensional joint distribution function with margins $F_1,\dots,F_n$. Then there exists an n-dimensional copula $C$ that satisfies the following equality for all $x = (x_1, \ldots, x_n) \in \mathbb{R}^n$: \begin{equation}
	    F(x_1,\ldots,x_n) = C(F_1(x_1),\ldots, F_n(x_n))
	\end{equation}Additionally, if all the marginals are continuous, then $C$ is unique. Conversely, if $C$ is a copula, and all the margins $F_1,\ldots,F_n$ are univariate distribution functions, then the function $F$ that satisfies the above equation is a joint distribution function with margins $F_1,\dots,F_n$.
\end{theorem}

\changeR{Using Equation~(6), we can now represent the joint density of multivariate random variables $\mathbf{X}=\{X_1,X_2, \ldots, X_n\}$ in terms of a copula function and the marginals. Thus, copula offers a simple, yet powerful approach to sample from the joint distribution of random variables with known marginals.} 

\subsection*{Copula for directional dependency} \label{section:copula}
The key idea behind accommodating directional dependency between two random variables, say $U$ and $V$, using copula is to construct an \emph{asymmetric} copula \cite{liebscher2008construction}. Formally, a bivariate asymmetric copula $C(u,v): [0,1]^{2} \rightarrow [0,1]$ is defined by any copula function that satisfies $C_{U|V}(u,v) \neq C_{V|U}(v,u)$, where $C_{U|V}(u,v) = \partial C(u,v)/\partial v$ and $C_{V|U}(v,u) = \partial C(v,u)/\partial u$. In this paper, we consider an asymmetric copula from the Rodriguez-Lallena and Ubeda-Flores family and is given as \cite{rodriguez2004new}: 
\begin{equation}
C(u,v,\phi) = uv+f(u)g(u) = uv + \vartheta uv (1-u)^{\alpha}(1-v)^{\beta}
\end{equation}
where $u,v\in [0,1]$, and $\phi = (\vartheta, \alpha, \beta)$, $\alpha>1, \beta>1$. The association parameter $\vartheta \in [-1, 1]$ measures the dependence between $u$ and $v$, while the asymmetry in the copula is captured by the parameters $\alpha$ and $\beta$. Given $n$ observations $\{(u_i, v_i)\}_{i=1}^{n}$, the maximum likelihood estimates (MLE) of $\alpha$ and $\beta$ are given as: 
\begin{align}
\begin{split}
\hat{\alpha} &=\frac{\sum_{i=1}^{n}((1-u_i)\log(1-u_i)-u_i)}{\sum_{i=1}^{n}u_i\log(1-u_i)}
\\
\hat{\beta} &=\frac{\sum_{i=1}^{n}((1-v_i)\log(1-v_i)-v_i)}{\sum_{i=1}^{n}v_i\log(1-v_i)}
\end{split}\label{eq:mles}
\end{align}
\changeR{Given $\alpha$ and $\beta$}, the admissibility bound for $\vartheta$ as shown by \cite{bairamov2001new} is given as:
\begin{equation}\label{admissibilityBound}
    -\min \left\{\left(\frac{\alpha + 1}{\alpha - 1}\right)^{\alpha - 1}, \left(\frac{\beta + 1}{\beta - 1}\right)^{\beta - 1}\right\} \leq \vartheta  \leq \min \left\{\left(\frac{\alpha + 1}{\alpha - 1}\right)^{\alpha - 1}, \left(\frac{\beta + 1}{\beta - 1}\right)^{\beta - 1}\right\}
\end{equation} 
Let $\rho_{v\rightarrow u}$ denote the degree of directional dependency of $U$ on $V$, with a higher value indicates a stronger directional dependency. Adopting the idea of \cite{sungur2005note}, $\rho_{v\rightarrow u}$ for the copula in Equation~(7) can be expressed as: 
\begin{equation}
\rho_{v\rightarrow u} = 12 \cdot   E\left[\left(r_{U|V}(u)\right)^2\right] -3 \label{DirecDep}
\end{equation}
where the expectation $E[\cdot]$ is taken with respect to the copula regression function $r_{U|V}(u)$ defined by $r_{U|V}(u) = 1-\int_{0}^{1}C_{U|V}(u,v)dv$. For the Rodriguez-Lallena and Ubeda-Flores copula, we can express Equation~(\ref{DirecDep}) in a closed-form \cite{sungur2005note,sungur2005some}: 
\begin{align}
\label{eq:directional_dependency}
 \rho_{v\rightarrow u}= \frac{3\vartheta^2\alpha^2\beta^2}{(2+\beta)^2(1+2\alpha)}\quad \text{and} \, \quad \rho_{u\rightarrow v}= \frac{3\vartheta^2\alpha^2\beta^2}{(2+\alpha)^2(1+2\beta)}  
\end{align}
\changeR{With this copula approach to model directional dependency, we now extend the Bayesian framework presented in Equations~(2)--(4) to incorporate directional dependency for multi-view clustering.}  

\section*{Copula-based multi-view clustering}
\changeR{In this section, we present our Dirichlet mixture model to incorporate directional dependence for multi-view clustering followed by the details of posterior inference for our model. }
\subsection*{Copula-based dirichlet mixture model}
The Dirichlet mixture model as presented in Equations~(\ref{eq:CDF_F_in_Dirichlet mixture model}) -- (\ref{eq:Dirichlet}) has been constructed for each data source, $m$, individually. Hence, there is no information borrowing between datasets. In what follows, we make use of the formula of directional dependency between a pair of data sets (see Equation~(\ref{eq:directional_dependency})), and subsequently utilize these quantities in clustering through the Dirichlet mixture model.

A hierarchical formulation for a copula-based Dirichlet mixture model based on the Rodriguez-Lallena and Ubeda-Flores copula is given as:
\begin{align}
X_{mi}|L_{mi}, \theta &\sim F(\theta_{mL_{mi}}); \quad
\theta_{mk} \sim G^{(0)}
\\
p(L_{1i}, \ldots L_{Mi}|\gamma, \rho) &\propto \prod_{m=1}^M \gamma_{m L_{mi}} \ \prod_{m=1}^{M} \prod_{k =1, k \neq m}^M~\left(1 + \rho_{m\rightarrow k} \ind_{\{m \rightarrow k\}} \ind_{\{L_{mi} = L_{ki}\}}\right)
\\
\gamma_{m1}, \gamma_{m2}, \ldots, \gamma_{mK} &\sim \text{Gamma}\left(\alpha_m/K,1\right)
\\
\rho_{m\rightarrow k} &\sim H_{m,k}
\end{align}
where $\text{Gamma}(a,b)$ denotes the gamma distribution with shape parameter $a$ and rate parameter $b$, and $\ind_{\{m \rightarrow k\}}$ is an indicator function such that $\ind_{\{m \rightarrow k\}}=1$ (otherwise zero) if there exists a known directional dependency from the $m$-th dataset towards the $k$-th dataset, while $\rho_{m\rightarrow k}$ is the degree of directional dependency associated with the two data types (see Equation~(\ref{eq:directional_dependency})). Notations $F$, $G^{(0)}$, and $\theta_{mk}$ are used in the same way  as used in the Dirichlet mixture model (Equations~\ref{eq:CDF_F_in_Dirichlet mixture model}) -- (\ref{eq:Dirichlet})). The weights $\gamma_{m L_{mi}}$ are derived from $\pi_{mk}$ as shown in Equation~(\ref{eq:MixtureModel}) such that $\pi_{mj} = \gamma_{mj} / \sum_{k = 1}^K \gamma_{mk}$. Finally, $H_{m,k}$ is the prior on the parameter $\rho_{m\rightarrow k}$ and it will be discussed in the following sections. 

\changeR{From the foregoing model, we note that: (a) the major distinction between the proposed model (Equations~(12) -- (15)) and the standard Dirichlet mixture model (Equations~(\ref{eq:CDF_F_in_Dirichlet mixture model}) -- (\ref{eq:Dirichlet})) is observed by the clustering allocation procedure induced by the latent variables $L_{1i}, L_{1n},\ldots, L_{Mi}$ and (b) the integration of directional dependency via $\rho_{u \rightarrow v}$ provides an advancement over the existing integrative models such as \cite{lock2013bayesian,kirk2012bayesian}.}

\subsection*{Directional dependence prior}\label{DepenPrior}
For each $i$, given two data types $u$ and $v$, let the notations $\halpha^{i}_{uv}$ and $\hbeta^{i}_{uv}$ denote the MLEs of the two parameters, $\alpha$ and $\beta$, of the Rodriguez-Lallena and Ubeda-Flores copula, given by Equation~(\ref{eq:mles}). Then after averaging each of the quantities over the $n$ observations, that is, $\halpha_{uv} = \frac{1}{n} \sum_{i = 1}^n \halpha_{uv}^{i}$ and $\hbeta_{uv} = \frac{1}{n} \sum_{i = 1}^n \hbeta_{uv}^{i}$, we can express the directional dependence of the $v$-th dataset on the $u$-th dataset as follows \cite{sungur2005some}:
\begin{align}
\label{eq:random_variable_directional_depedency}
\rho_{u\rightarrow v} = \frac{3\vartheta^2_{uv} \halpha_{uv}^2 \hbeta_{uv}^2}{(2+\halpha_{uv})^2(1+2\hbeta_{uv})}  
\end{align}
where $\vartheta_{uv}$ is a measure of association between the random variables $U$ and $V$. Note that $\rho_{u\rightarrow v}$ in Equation~(\ref{eq:random_variable_directional_depedency}) is a \emph{copula-based random variable}: first, two quantities $\halpha_{uv}$ and $\hbeta_{uv}$ are driven from a directional copula; and second, $\vartheta_{uv}$ is the stochastic part, rendering $\rho_{u\rightarrow v}$ as a random quantity.

\changeR{In this work, we consider a Gaussian distribution for the random variable $\vartheta_{uv}$, given by $\vartheta_{uv} \sim N\left(0, b_{uv}^2/9\right)$, where the $b_{uv}$ is obtained by the admissibility bound presented in Equation~\eqref{admissibilityBound} as}: 
\begin{equation} 
b_{uv} = \underset{i\in \{1,\ldots,n\}}{\min}\left\{\left( \frac{\halpha_{uv}^{i} + 1}{\halpha_{uv}^{i} - 1}\right)^{\halpha_{uv}^{i} - 1}, \, \left( \frac{\hbeta_{uv}^{i} + 1}{\hbeta_{uv}^{i} - 1}\right)^{\hbeta_{uv}^{i} - 1} \right\}
\end{equation}

The motivation behind this choice of $b_{uv}$ is as follows: as a Gaussian random variable, $\vartheta_{uv}$ has a variance of $b_{uv}^2/9$ such that $\text{P}(|\vartheta_{uv}| \leq b_{uv}) = 0.997$. This implies that $-b_{uv} \leq \vartheta_{uv} \leq b_{uv}$ holds with a very high probability. {Since $-b_{uv} \leq \vartheta_{uv} \leq b_{uv}$ holds with a very high probability, the posterior updates are all conjugate updates without the added computational burdens of truncated distributions.} Moreover, as $\vartheta_{uv}$ is normally distributed, it is easy to see that $\rho_{u\rightarrow v} \propto \vartheta_{uv}^2$ and follows a gamma distribution. In particular, \begin{equation}
    \rho_{u\rightarrow v} \sim \text{Gamma}\left(\frac{1}{2},  \frac{4(2+\halpha_{uv})^2(1+2\hbeta_{uv})}{2\halpha_{uv}^2\hbeta_{uv}^2(b_{uv}/3)^2}\right)
\end{equation}This defines our prior $H_{u,v}$ on $\rho_{u\rightarrow v}$.

\subsection*{Posterior updates}
\changeR{In this section, we present a general Bayesian framework to estimate the posterior updates using a Gibbs sampling approach. For this, we follow the posterior inference, as presented in \cite{kirk2012bayesian}. We begin by referring to our general model in Equation~(13). We first obtain the joint density of the latent allocation variable $L_{mi}$'s by defining a normalizing constant $Z$ as: 
 \begin{equation}
     Z = \sum_{j_1 = 1}^K \cdots \sum_{j_M = 1}^K \left[ \prod_{k = 1}^M \gamma_{kj_k} \prod_{k = 1}^M\prod_{\ell = 1, \ell \neq k}^M \left(1 + \rho_{k \rightarrow \ell} \ind_{\{k \rightarrow \ell\}} \ind_{\{j_k = j_{\ell}\}} \right)\right]
 \end{equation}such that the joint density for $N$ objects is given as:
 \begin{equation}
     p(\{L_{1i}, \ldots L_{Mi}\}_{i=1}^{N})  = \frac{1}{Z^N}\prod_{i=1}^{N} \left[ \prod_{k = 1}^M \gamma_{kj_k} \prod_{k = 1}^M\prod_{\ell = 1, \ell \neq k}^M \left(1 + \rho_{k \rightarrow \ell} \ind_{\{k \rightarrow \ell\}} \ind_{\{j_k = j_{\ell}\}} \right)\right]
 \end{equation}
 \noindent Following \cite{nieto2004normalized}, we define the following joint density function using a strategic latent variable $\xi$ to provide the basis for a fully Bayesian framework: 
 \begin{equation}
     p(\{L_{1i}, \ldots L_{Mi}\}_{i=1}^{N}, \xi)  = \frac{\xi^{N-1}\exp(\xi Z)}{(N-1)!}\times p(\{L_{1i}, \ldots L_{Mi}\}_{i=1}^{N}) 
 \end{equation}}such that the conditional distribution for the latent variable is given as: \begin{equation}
     \xi|- \sim \text{Gamma}(N, Z)
 \end{equation} 
\noindent Subsequently, we note that the conditional on $\gamma_{mj_m}$ is given as:
\begin{equation}
    \gamma_{mj_m}|- \sim \text{Gamma}(a_{\gamma}, b_{\gamma})
\end{equation} 
\begin{equation}
    a_{\gamma} = \sum_{i = 1}^{N}\ind\{L_{mi} = j_m\} + \frac{\alpha_M}{K}
\end{equation} 
\begin{adjustwidth}{-2cm}{0cm}
\begin{equation}
b_{\gamma} = \xi \sum_{j_1 = 1}^{K} \cdots \sum_{j_{m-1} = 1}^{K} \sum_{j_{m+1} = 1}^{K} \cdots \sum_{j_M = 1}^{K} \left[ \prod_{k = 1, k \neq m}^{M} \gamma_{kj_k} \prod_{k=1}^{M} \prod_{\ell = k, \ell \neq k}^{M} \left( 1 + \rho_{k \rightarrow \ell} \ind_{\{k \rightarrow \ell\}} \ind_{\{j_k = j_{\ell}\}} \right)\right]  + 1
\end{equation}
\end{adjustwidth}

\noindent In a similar fashion, the conditional on $\rho_{m\rightarrow p}$ may be deduced as: 
\begin{equation}
    \rho_{m \rightarrow p}| -\sim \text{Gamma}(a_{\rho}, b_{\rho})
\end{equation}
\begin{equation}
    a_{\rho} = \sum_{i = 1}^N \ind\{L_{mi} = L_{pi}\} + \frac{1}{2}
\end{equation}

\begin{adjustwidth}{-2cm}{0cm}
\begin{equation}
\begin{split}
    b_{\rho} = \xi \sum_{j_1 = 1}^K \cdots \sum_{\substack{j_{m-1} = 1 \\ j_{m+1} = 1}}^K  & \cdots \sum_{\substack{j_{p-1}=1 \\ j_{p+1} = 1}}^K \cdots \sum_{j_M = 1}^K \left[ \prod_{k = 1}^M \gamma_{kj_k} \prod_{k = 1}^{M}\prod_{\substack{\ell = 1\\ \ell \neq k, \ell \neq p}}^{M} \left( 1 + \rho_{k \rightarrow \ell}\ind_{\{k \rightarrow \ell\}} \ind_{\{ L_{ki} = L_{\ell i}\}} \right) \right. \times \\ & \left.  \prod_{\substack{k = 1 \\ k \neq m, k \neq p}}^{M}  \left( 1 + \rho_{k \rightarrow p}\ind_{\{k \rightarrow p\}} \ind_{\{ L_{ki} = L_{pi}\}} \right) \right] + \frac{4(2+\halpha_{mp})^2(1+2\hbeta_{mp})}{2\halpha_{mp}^2\hbeta_{mp}^2(b_{mp}/3)^2}
\end{split}
\end{equation}
\end{adjustwidth}

\noindent Finally, the conditional distribution for the latent variables $L_{mi}$ is given as:
\begin{adjustwidth}{-2cm}{0cm}
\begin{equation}
\begin{split}
p(L_{mi} = c|\gamma, &\rho, x_{mi}, x_{m,-i}^c, L_{-m,i}, L_{m,-i}) = b_L \gamma_{mc} \prod_{k=1}^{m-1}\left(1 + \rho_{k \rightarrow m} \ind_{\{k \rightarrow m\}}\ind_{\{L_{ki} = L_{mi}\}}\right) \times \\ & \prod_{k = m+1}^{M}\left( 1 + \rho_{m \rightarrow k} \ind_{\{m \rightarrow k\}} \ind_{\{L_{mi} = L_{ki}\}} \right) \times \int f_m(x_{mi}, x^{c}_{m, -i}|\theta_m)g_m^{(0)} (\theta_m)d\theta_m
\end{split}  
\end{equation}
\end{adjustwidth}where $x_{mi}$ is observation $i$ for data type  $m$, $x^c_{m,-i}$ are all the observations not including $x_{mi}$ in data type $m$ associated with component $c$, $L_{m, -i}$ are all the $L_{mj}$ such that $i \neq j$, $L_{-m, i}$ are all the $L_{ki}$ such that $k \neq m$, and $b_L$ is a normalizing constant that ensures $\sum_{c = 1}^K p(L_{mi} = c|\gamma, \rho, x_{mi}, x_{m,-i}^c, L_{-m,i}, L_{m,-i}) = 1$. {Please note that a latent variable $\xi$ has been added to help with the computational efficiency and that the updates on $\theta_{m}$ will depend on the choice of $f_m(\cdot)$ and $G^{(0)}$. } \changeR{The posterior updates are based on pairwise directional dependence. To obtain the consensus/global clustering, we  leverage the fundamental idea of the central dogma, i.e., the directional dependence in genomic data. Based on prior studies~\cite{qin2019identifying}, we note that the more remote an omics level or data source is from a physiological trait, the smaller the magnitude of their correlation is. Since protein data is closest to the physiological trait (i.e., cancer type), we deem the clustering result of the protein as the final or global clustering.} 

\section*{Results}
To demonstrate the effectiveness of \changeR{incorporating} directional dependencies in multi-view clustering, we consider both simulated and real-world examples. We also compare the performance of our approach with competing methods such as \cite{lock2013bayesian} and \cite{kirk2012bayesian}. 

\subsection*{Simulated Data}
For the simulation experiment, we consider two data types, $U$ and $V$, each generated from a univariate Gaussian mixture model with two components. The corresponding means and standard deviations of the mixture model are $0,3$ and $1,0.5$, respectively, and the directional dependency is captured by an asymmetric Tawn copula given as: 

\begin{equation}
    C(u,v) = \exp\left(\log(u) + \log(v) A\left(\frac{\log(v)}{\log(u)\log(v)}\right)\right)
\end{equation} where $A(.)$ is called the Pickands dependence. For the Tawn copula, the Pickands function is given as: 
\begin{equation}
A(t) = (1-\psi_1)(1-t) + (1-\psi_2)t + [(\psi_1(1-t))^{\theta} + (\psi_2t)^{\theta}]^{1/\theta}
\end{equation} In this case, we consider the Tawn copula of Type 1 for which $\psi_2 = 1$. For the current simulation study, we set the values of $\psi_1$ and $\theta$ to $0.5$ and $30$, respectively, and the direction of dependence is set from $V$ to $U$. Details on the Tawn copula can be found in \cite{kraus2017d}. For each data type, we generate 500 data points, and the measure of directional dependence (from $V$ to $U$) is estimated from Equation~(11). For the Tawn copula constructed above, the directional dependence is equal to 1.73. Note that the dependence in the opposite direction was obtained to be -1.02, indicating no notable dependence.

\changeR{We test the performance of the proposed methodology using three different scenarios:
\begin{enumerate}[leftmargin=0pt]
    \item[](i) when the true directionality is used, i.e., $\ind_{\{V \rightarrow U\}}$,
    \item[](ii) when there is no direction of dependence, and
    \item[](iii) when the direction of dependence is reversed, i.e., $\ind_{\{U \rightarrow V\}}$.
\end{enumerate}} 

\begin{center}
\end{center}
\begin{figure}[h]
    \centering
    \includegraphics[width = 0.85\textwidth]{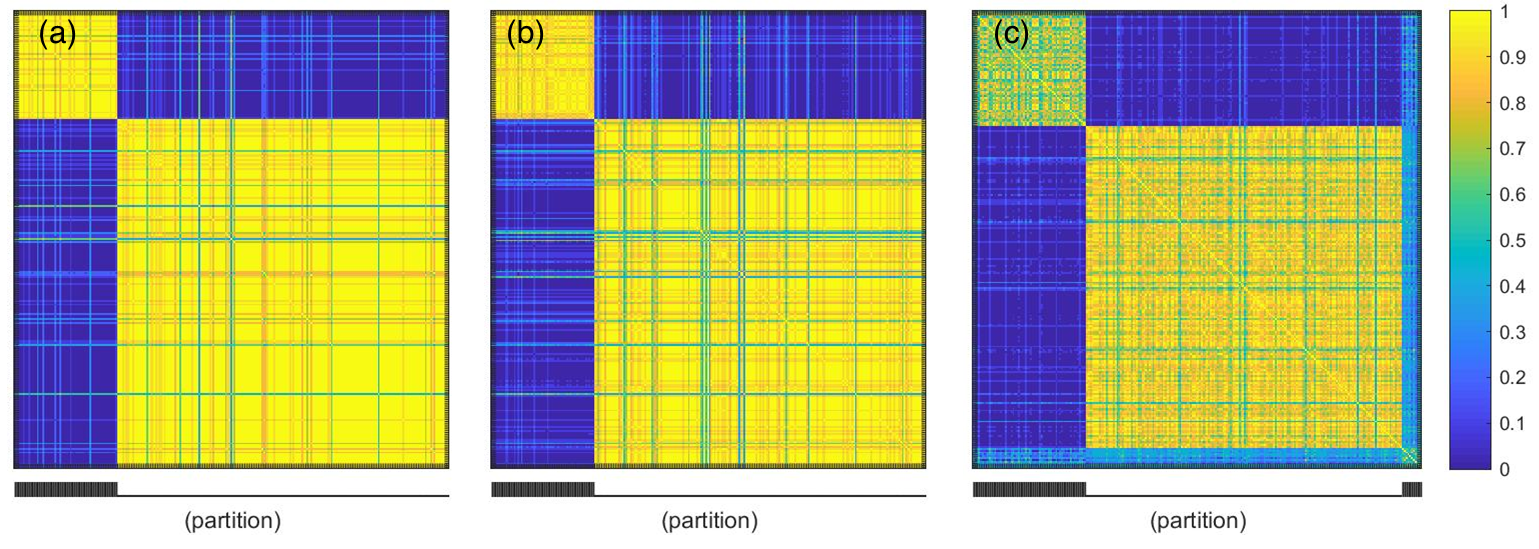}
    \caption{{ \textbf{Similarity matrix for the simulation study}: (a)  the true direction of dependence is considered, (b) no directional dependence is considered, (c) the direction of dependence is reversed. The colormaps show the posterior probability of samples $i$ and $j$ to belong to the same cluster.}}
    \label{fig:summatrix}
\end{figure}

In the first case, the method is able to correctly predict the 2 clusters and is shown in the joint similarity matrix in Fig.~\ref{fig:summatrix}(a). The similarity matrix displays the posterior probability of samples $i$ and $j$ to belong to the same cluster  (see \cite{kirk2012bayesian} for details). The overall accuracy of clustering for this case is 97.8\%. The clustering results corresponding to cases (ii) and (iii) are presented in the joint similarity matrix shown in Figs.~\ref{fig:summatrix} (b) and (c). \changeR{We note that the clustering performance is affected in both the cases, but more in case (iii), where we observe three different clusters as opposed to two clusters as in cases (i) and (ii). The intuition behind the significantly worse performance of case (iii) as compared to cases (i) and (ii) is that in case (iii), we wrongly reverse the directional dependence between the data types. In contrast, case (ii) is more reminiscent of the integrative analysis presented by \cite{kirk2012bayesian, lock2013bayesian}}. The clustering outcome is affected since the underlying models consider dependence without any directionality. Additionally, in case (ii), where no directionality is considered, we do get two clusters. However, the clustering accuracy is 97.2\%, lower than that of the first case.  

\changeR{In addition to studying the effect of causal relations on the clustering performance, we also investigate the effect of different copulas and sample size. We first look at the effect of different copulas. In this simulation study, we refer to three different copulas: Tawn Type 1 (TT1), Tawn Type 2 (TT2), and BB1 copula. See \cite{kraus2017d} for the functional form and dependence structure. Note that by varying the copula model, we essentially modify the dependence structure. Results obtained from 10 simulation runs show that clustering accuracy for TT1, TT2, and BB1 copulas for each of the aforementioned cases (case (i), (ii), and (iii)) are $(0.97,0.96,0.84)$, $(0.95,0.93,0.94)$, and $(0.97,0.92,0.94)$, respectively. The average number of clusters were recorded as $(2,2,3.67)$, $(2,3,3.4)$, and $(2,3,2)$. We note that by incorporating the directional dependency, we are able to consistently achieve a higher accuracy as well as identify the two clusters. For case (ii), i.e., integrative clustering without any direction of dependence achieves marginally lower accuracy but fails to identify the two clusters correctly. Case (iii) performs worse both in terms of accuracy and the number of clusters. }

\changeR{Next, we look at the effect of the sample size. For this, we fix our copula to Tawn Type 1 and obtain the results for three different sample sizes: $250, 500, 750$. Corresponding to these sample sizes, the clustering accuracy was noted as $(0.97,0.96,0.84)$, $(0.97,0.97,0.87)$, $(0.97, 0.96,0.64)$. In addition, the average number of clusters were obtained as $(2,2,3.67)$, $(2,2,3)$, and $(2,2 ,5)$. We again note that the proposed methodology performs better as compared to the case when no directionality is considered and when the direction is reversed.} 

\changeR{We also note that the computational complexity scales linearly with the sample size. In fact, the computational complexity of the present method is directly proportional to the number of data sources $M$, the number of clusters $K$, and the number of genes $N$. So the algorithm scales as $O(NMK)$. For sample sizes of 250, 500, and 750, the algorithm converges in 167.2 sec, 364.5 sec, and 520.16 sec when running in parallel. Convergence of the method can also be argued from the standpoint of clustering results, i.e., as sufficient data is made available, the clusters estimated from the method will closely resemble the true clusters \cite{lock2013bayesian}.}

\subsection*{TCGA Breast Cancer Data}
For \changeR{the application of our model to} a real-world dataset, we considered the breast cancer tumor samples from TCGA \changeR{program that consists of multi-source genomic data for a common set of patients}. \changeR{Since breast cancer is a heterogeneous disease and can be effectively used in the case studies for clustering models. This dataset has become a benchmark dataset used in several multi-view clustering studies, such as \cite{lock2013bayesian}}. This dataset is available for  download from the web portal of TCGA (\url{https://www.cancer.gov/tcga}) and contains a common set of 348 breast cancer tumor samples (i.e., $N = 348$) and four distinct data types (i.e., $M=4$):
\begin{itemize}
\item RNA gene expression (GE) data for \changeR{645} genes.
\item DNA methylation (ME) data for \changeR{574} probes.
\item miRNA expression (miRNA) data for \changeR{423} miRNAs.
\item Reverse phase protein array (RPPA) data for 171 proteins.
\end{itemize}
\changeR{We chose the 171 genes, probes, and miRNAs corresponding to the 171 existing proteins in the other three data types. These 171 genes (or their product proteins) are carefully chosen to contain the genes such as PIK3CA, PTEN, AKT1, TP53, GATA3, CDH1, RB1, MLL3, MAP3K1, and CDKN1B that are well-known to be important for classification of breast cancer subtypes \cite{cancer2012comprehensive}}. 

\begin{figure}[!htb]
\centering
\includegraphics[width = 0.85\textwidth]{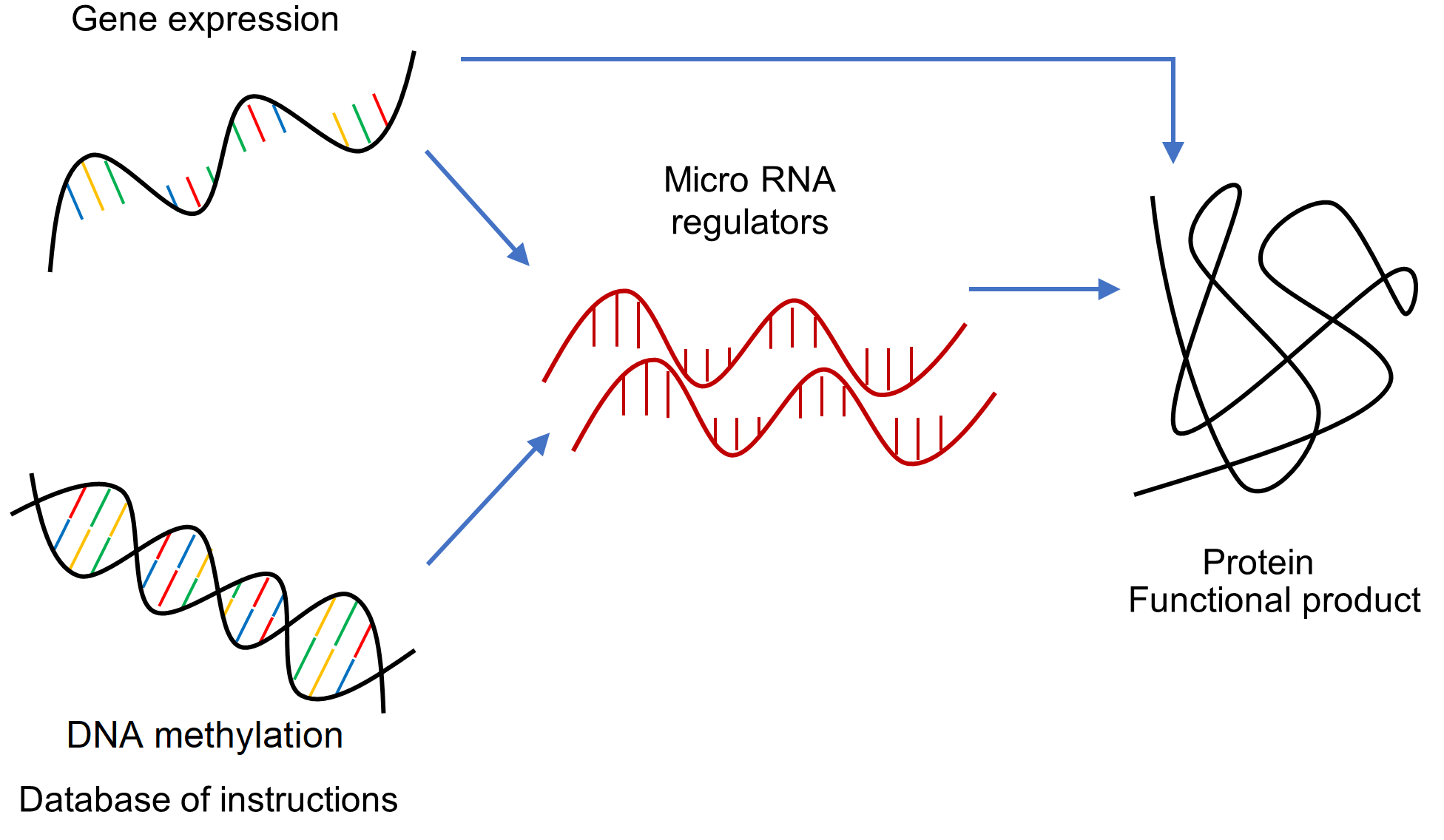}
\label{fig:Relationship}
\caption{\textbf{The central dogma of molecular biology}. Directional dependencies among biological components}
\label{tt}
\end{figure}

It is known that these four data types manifest differently, but at the same time, are highly related in that they are directionally dependent~\cite{crick1958protein, crick1970central}. This directional dependence can be determined by the central dogma of molecular biology, where the transcription and translation process determines the direction of dependence between DNA to RNA gene expression and from RNA gene expression to protein, respectively. \red{Figure~\ref{tt} shows the four data types used in this study, along with the direction of dependence between them. We consider three pathways for directional dependence. The first dependence pathway is from RNA gene expression to protein. This is the fundamental relationship, as explained by the central dogma. The second dependence pathway is from RNA gene expression to protein via miRNA (or microRNAs). MiRNAs are single-stranded RNAs that exert their regulatory action by binding RNAs gene expression and preventing their translation into proteins. The last pathway is from DNA methylation to protein via miRNAs. This dependence is primarily based on recent studies that have shown the influence of DNA methylation directly on the expression of miRNAs \cite{glaich2019dna}. Note that these relationships are not exhaustive and additional experiments will be needed to fully model the pathways that influence the production of protein from DNA.} 

\begin{center}
\end{center}

\changeR{As discussed in the foregoing, the central dogma explains that this transfer of information has a pre-specified direction, for example, the transfer of information is from gene to protein and not from protein to protein or protein to gene \cite{weber2006central}}.
From a statistical perspective, both \emph{transcription} and \emph{translation} might be designed in terms of directional dependency in our copula model because opposite \changeR{dependencies (i.e., Protein to RNA and RNA to DNA) may not exist}. Considering the two directions are \emph{a priori} known, we can design them by providing deterministic directional indicators $\{ k \rightarrow p\}$ for each process where $k$ and $p$ are indices for the corresponding data types. Also, the corresponding strengths of directional dependencies are quantified by $\rho_{k \rightarrow p}$. \changeR{From a numerical perspective, we note that providing a deterministic directional indicator into our model reduces the computational burden in the summation of $b_{\rho}$, and therefore, contributes to the computation speed}.

Since we use the copula at the data level, not at the latent level, matching the data dimensions for each data type is necessary ($D_{1} =D_{2}=D_{3}=D_{4} =171 $). \changeR{This is because we modeled the dependence between the data types through the directional relationship between their features (such as genes, RNAs, proteins, etc.)}. These four data types are measured on different platforms and represent different biological components. However, they all represent genomic data for the same sample set, and it is reasonable to expect shared structure while considering directional dependencies at hand~\cite{lock2013joint}.

As explained in the previous section, we are clustering samples based on four data types: gene expression, DNA methylation, microRNA, and RPPA for breast cancer from the TCGA data. Prior studies using this data have found that the total number of clusters can vary from two \cite{duan2013metasignatures} to 10 \cite{curtis2012genomic}. \changeR{However, mainly four prominent} subtypes have been identified based on multi-source consensus clustering of the TCGA data as Basal, Luminal A, Luminal B, and HER2 \cite{cancer2012comprehensive}. We incorporated our prior biological knowledge, i.e., the directional dependence based on the central dogma {\cite{crick1970central,weber2006central}}, into our integrative clustering algorithm. Since proteins are the final outcome,  we consider the consensus (or final) clustering to be the protein clusters, which summarize all the information from the other three datasets inside itself. 

To initialize the Bayesian posterior update algorithm, we take advantage of our finite Dirichlet mixture model to define the number of clusters $(K)$. Although our model considers a finite mixture model, it is equivalent to a Dirichlet process mixture model when $N\rightarrow \infty$ and therefore, $K$ specifies an upper bound on the number of clusters present in the data. Authors in \cite{rousseau2011asymptotic} argue that if the number of clusters specified by $K$ is sufficiently large, the posterior updates can automatically determine the true number of clusters present in the data. Based on this analogy, \cite{kirk2012bayesian} suggest that a \changeR{pragmatic choice for the upper bound on the number of} cluster to avoid the computational burden is $K =\left \lceil{N/2}\right \rceil $. \changeR{From our experimental studies, we note that even if this upper bound is as high as 500}, our algorithm correctly predicts the number of clusters to be four, similar to the sub-typing in TCGA. \changeR{This shows that the clustering results are not contingent on the choice of $K$.} 


\begin{table}[!b]
        \caption{\textbf{Confusion matrix for the clustering assignment}. The results are shown for the clustering of four cancer data types: Her2, Basal, Lum A, and Lum B. }
    \begin{tabular}{|c|c|c|c|c|}
    \hline
    \backslashbox{True cluster}{Estimated cluster}
        & 1 & 2 & 3 & 4\\
        \hline
        \hline
        Her2 & \red{5} & \red{4} &\red{22} & \red{8}\\
        \hline
        Basal & \red{2} & \red{5} & \red{48} & \red{17}\\
        \hline
        Lum A & \red{41} & \red{103} & \red{2} & \red{15}\\
        \hline
        Lum B & \red{34} & \red{22} & \red{5} & \red{15} \\
        \hline
    \end{tabular}
    \label{tab:ours}
\end{table}

\begin{figure}[h]
    \includegraphics[width=\textwidth]{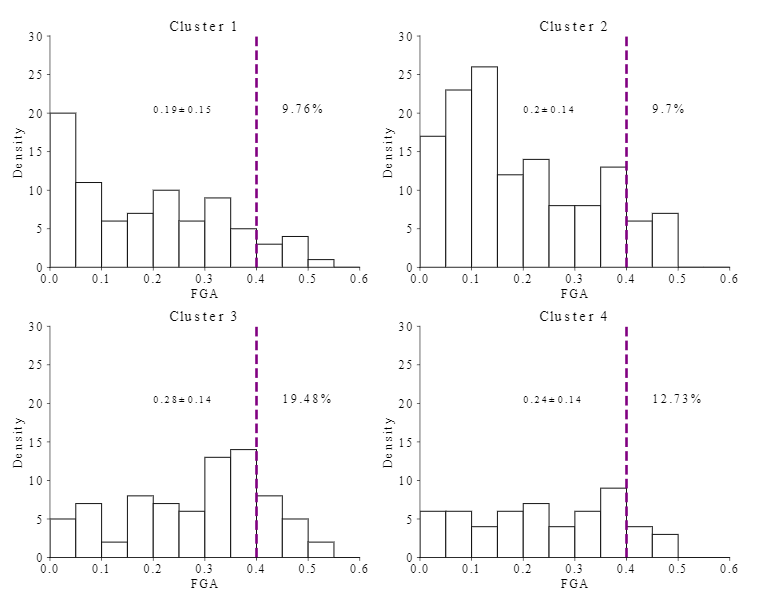}
    \caption{\textbf{Distribution of FGA across four clusters of breast cancer}. \changeR{Dashed line represents the genes with high FGA values.}}
    \label{fig:AFG}
\end{figure}

Since copy number variation or the \changeR{focal amplification/deletion of a region of gene}, is associated with breast cancer risk and prognosis \cite{kumaran2017germline,fan2019benchmarking,mallory2020assessing,knijnenburg2018genomic,edrisi2019combinatorial}, we calculate the fraction of the genome altered (FGA) as a measure of copy number activity as described in the Supplementary Section VII of~\cite{cancer2012comprehensive} (with copy number level threshold $T=0.15$) for each cluster. Our results are summarized in Table \ref{tab:ours} and visualized in Fig.~\ref{fig:AFG}. The TCGA breast cancer subtypes and our clusters have different structures, but they are non-independent based on the Chi-squared test of independence (p-value $<$ 0.0001). Clusters 1 is mostly a combination of Luminal like breast cancer subtypes (Luminal A and Luminal B), which are similar to each other with average FGA of \red{$0.19\pm 0.15$} and almost \red{10$\%$} of their samples have high FGA of more than 0.4. Cluster 2 is mostly Luminal like breast cancer B with higher FGA (\red{$0.2\pm 0.14$}). Moreover, \red{cluster 2 contains similar high FGA in comparison to cluster 1 with almost 10$\%$}. Cluster 3 contains more of HER2 and Basal subtypes, which are more similar to each other with the highest FGA (\red{$0.28\pm 0.14$}) and very high FGA (approximately \red{19$\%$}). Even though cluster 4 is more spread over the four known subtypes (Her2, Basal, Luminal A and Luminal B), they include samples with high FGA (\red{$0.24\pm 0.14$}) and lower standard deviation compared to cluster 2. Moreover, cluster 4 is second in having high FGA samples with almost \red{13$\%$}.

As mentioned in the foregoing, two other state-of-the-art algorithms for integrative clustering exists: Bayesian Consensus Clustering (BCC) \cite{lock2013bayesian} and Multiple Dataset Integration (MDI) \cite{kirk2012bayesian}. However, it is not feasible to compare our results with MDI as it provides separate clustering for each data type as opposed to a consensus or global clustering. To compare the performance of our method with BCC, we use the Rand index as the number of clusters reported by BCC is three, while four clusters exist in the TCGA dataset. Essentially, the Rand index measures the level of similarity between two clustering methods without employing the data labels \cite{rand1971objective}. When one of the clustering methods is the ground truth, it essentially measures the proportion of the correct allocations. We note that the Rand index of BCC is 0.68, and for the proposed method, it is \red{0.70.  Not only BCC performs marginally worse in terms of correctly labeling the datasets, but also the algorithm needs to know the true number of clusters beforehand, which is a major limitation.} The authors developed a heuristic approach to calculate the number of clusters as a pre-processing step that suggested only three clusters as opposed to four true clusters predefined in the TCGA dataset. In comparison, our integrated analysis method correctly identified the four clusters out of the maximum number 174 potential clusters ($K =\left \lceil{N/2}\right \rceil $).

\begin{center}
\end{center}

\section*{Conclusion}
It is known that the genomics data types collected from multiple sources are often related, and their integrated analysis can significantly improve the downstream analysis such as the clustering outcome. \changeR{Several studies have been proposed in the literature for integrated analysis of multi-view data that attempts to capture the association or dependence between different data types. Nonetheless, the dependence between real-world data types often have many added level of complexity due to the underlying natural phenomena.} This is often true in genomics, where, underlined by the central dogma, the data types are not only dependent but directionally dependent. We utilized this domain knowledge and proposed a novel method for multi-view clustering by incorporating the pre-specified directional dependence between the genomic data types using a copula model. The use of copulas to model the directional dependence provides a robust and versatile tool to capture the directional dependence in joint behavior. The application of the proposed method on synthetic as well as real datasets demonstrates its efficacy. Most importantly, we believe that capturing directional dependence instead of simple dependence can provide an added understanding of the underlying process. 

\changeR{With the groundwork of directionally dependent multi-view clustering in this work, several improvements can be made over the proposed model. Firstly, we can utilize spike-slab priors\cite{ishwaran2005spike,cui2012spike,rockova2020dynamic} for feature selection while performing directional clustering in high-dimensional TCGA applications. \changeR{Secondly, an approach to deal with a different number of features in each data type can add more flexibility to the proposed model}. Thirdly, we may incorporate hidden Markov models (HMM) in our proposed approach for modeling longitudinal clustering where the class type (or process state) of a patient is a hidden variable, and the multiple data sources are the observed variables. \cite{helske2017mixture,altman2007mixed,maruotti2011mixed}.} Finally, more rigorous and in-depth comparative analysis between the dependence and directional dependence seeking multi-view clustering with several datasets are needed. 

\section*{Acknowledgement}
Research reported in this publication was partially supported by the National Cancer Institute of the National Institutes of Health under award number R01CA194391, NSF grants numbers NSF CCF-1934904, NSF IIS-1741173.


%

\bibliographystyle{plos2015}

{\footnotesize\bibliography{ref.bib}}
\end{document}